\documentclass{svjour3.arxiv}                     %
\smartqed  %
\usepackage{graphicx}
\usepackage{amssymb}
\usepackage{url}

\newcommand{\eq}{\begin{equation}}
\newcommand{\en}{\end{equation}}

\newcommand{\ed}{\mbox{$ \ \stackrel{d}{=}$ }}

\newcommand{\br}{B^{\mbox{$\scriptstyle{\rm br}$}}}

\begin{document}

\title{Efficient computation of the cdf of the maximal difference between Brownian bridge and its concave majorant}

\author{Fadoua Balabdaoui \and Karim Filali}

\institute{F. Balabdaoui \at
              Universit\'e Paris-Dauphine \\
              Tel.: +33-1-44-05-48-83\\
              Fax: ++33-1-44-05-45-99\\
              \email{fadoua@ceremade.dauphine.fr}           %
           \and
           K. Filali \at
              Yahoo Labs and University of Washington \\
               \email{karim@cs.washington.edu}
}

\date{February 26, 2010}

\maketitle

\begin{abstract}
In this paper, we describe two computational methods for calculating the cumulative distribution function and the upper quantiles of the maximal difference between a Brownian bridge and its concave majorant. The first method has two different variants that are both based on a Monte Carlo approach,  whereas the second uses the Gaver-Stehfest (GS) algorithm for numerical inversion of Laplace transform.  If the former method is straightforward to implement, it is very much outperformed by the GS algorithm, which provides a very accurate approximation of the cumulative distribution as well as its upper quantiles.  Our numerical work has a direct application in statistics: the maximal difference between a Brownian bridge and its concave majorant arises in connection with a nonparametric test for monotonicity of a density or regression curve on $[0,1]$. Our results can be used to construct very accurate rejection region for this test at a given asymptotic level.

\end{abstract}

\section{Introduction}
\label{intro}

Consider the regression model $Y_i = f_0(t_i) + \epsilon_i$, where $t_i = i/n$ for $i=1, \cdots, n$, and conditionally on the regressors $t_i$'s, $\epsilon_1, \cdots, \epsilon_n$ are i.i.d. $\sim (0, \sigma^2_0)$ with $0 < \sigma_0 < \infty$. Suppose that we are interested in knowing whether the true regression curve $f_0$ is nondecreasing on some sub-interval of $[0,1]$. \cite{durot03} considered the nonparametric test based on the maximum difference between the cumulative sum diagram of the observations and its concave majorant, multiplied by $\sqrt{n}$ and divided by any consistent estimator of $\sigma_0$.

The intuition behind this test is as follows: Under the null hypothesis that $f$ is decreasing on $[0,1]$, the function $\int_0^x f_0(t)dt$, $ x \in [0,1]$, is concave, and hence the cumulative sum diagram of the data must be \lq\lq very close\rq\rq \ to its concave majorant as $n \to \infty$.  \cite{durot03} showed that asymptotically, the Type I error of the test attains its maximum when $f_0$ is constant on $[0,1]$. Then, under this least favorable case, the test statistic converges weakly to the maximum difference between a standard Brownian motion on $[0,1]$ starting at 0 and its concave majorant.  If $(B(t), 0 \le t \le 1)$, $B(0) = 0$ denote a standard Brownian motion on $[0,1]$ starting at 0 and $\widehat B$ its concave majorant on $[0,1]$, then Durot's test statistic converges weakly to
\begin{eqnarray}\label{M}
M \ed \sup_{t \in [0,1]} \left( \widehat B(t) - B(t)\right).
\end{eqnarray}
A similar testing problem for densities was considered by \cite{kuliklop08}. The test can be based on the maximum difference between the empirical distribution and its concave majorant, multiplied by $\sqrt{n}$.  When the true density is uniform on $[0,1]$, this maximum difference converges weakly to the distribution of $M$ as in the regression setting above.

As proved in Proposition 4 (iii) in \cite{balabdpitman09}, one interesting property of the distribution of $M$ is that we can replace $B$ in (\ref{M}) by a standard Brownian \textit{bridge}; i.e, the distribution of $M$ is also that of the maximum difference between a standard Brownian bridge and its concave majorant. Furthermore, the random variable $M$ can be given under a more useful form. Let $M_3$ denote the maximum of a Brownian excursion and $(M_{3,1}, M_{3,2}, \cdots) $ an infinite sequence of independent random variables distributed as $M_3$. If $(U_1, U_2, \cdots )$ is an infinite sequence of independent uniform random variables on $[0,1]$ and $(L_1, L_2, \cdots) $ the corresponding uniform stick-breaking process; i.e.,
$$
L_1:= U_1, \ L_2:= (1-U_1) U_2, \ L_3:= (1-U_1)(1-U_2)U_3, \cdots
$$
then \cite{balabdpitman09} proved in Theorem 1 that
\begin{eqnarray}\label{MainTheo}
M \ed \max_j \sqrt{L_j} M_{3,j}.
\end{eqnarray}
Absolute continuity of $M$ is an immediate corollary of (\ref{MainTheo}). Two other corollaries will follow from the same equality in distribution giving formulae for $F$, the cdf of $M$. If $F_3$ is the cdf of $M_3$, then
\begin{eqnarray}\label{ExpMonteCarlo}
F(x) = E\left[\prod_j F_3\left( \frac{x}{\sqrt{L_j}} \right) \right], \ \ \forall \ x > 0
\end{eqnarray}
The expectation in the formula above is taken with respect to the stick-breaking process, and $F_3$ is known to be given by
$$
F_3(x) = 1 - 2\sum_{n=1}^\infty (4n^2 x^2 -1) \exp(-2 n^2 x^2), \ \ \forall \ x > 0,
$$
see e.g. \cite{kennedy76} and \cite{bianeetal01}.

A second formula, which follows from Proposition 7 and Theorem 8 of \cite{balabdpitman09}, gives $F$ as a function of the inverse of a Laplace transform. Let $K_m$ denote the modified Bessel function of the second kind and order $m \in \mathbb N$, and $G$ the function defined by
\begin{eqnarray}\label{G}
G(t) = \prod_{j=1}^\infty \exp\left( -4 \left( 2 \sqrt{2} t n K_1(2 \sqrt{2} n t) - K_0(2 \sqrt{2}n t) \right) \right), \ \ t > 0.
\end{eqnarray}
Then,
\begin{eqnarray}\label{ExpInvLT}
F(x) = L^{-1} \left( \frac{G(\sqrt{t})}{t} \right)\left(\frac{1}{x^2} \right), \ \ x > 0
\end{eqnarray}
where $L^{-1}h(z)$ denotes the value of the inverse of Laplace transform of $h$ at $z$.

We describe in Section 2 and 3 the implementation of two variants based on a Monte Carlo approach and a Gaver-Stehfest algorithm for approximating the inverse of Laplace transform. If the Monte Carlo (MC) methods are easy to implement, they both require a very large number of simulations in order to obtain the same precision as the deterministic Gaver-Stehfest (GS) algorithm. For $x \ge 0.33$,  GS is able to approximate very accurately the cumulative distribution of $M$ at $x$ using a multiple precision library. For values $x$ below what it seems to be a cut-off point for both methods, it is difficult to get a precise approximation for the distribution of $M$. This problem does not affect the calculation of the upper quantiles which is one of the main motivations of the work. Although the MC approach is not as efficient as the GS algorithm, it seemed natural to describe it in the sequel. We only report the numerical results of the GS algorithm, however. Tables \ref{DisFun1}, \ref{DisFun2} and \ref{DisFun3} below give approximated values of the distribution function of $M$ on a grid of real numbers $x$ such that $0.33 \le x \le 2.54$ with a regular mesh equal to 0.01. The approximation was performed with a precision ensuring up to 60 significant digits.  A table of quantiles of order $p \in \{0.90, 0.91, \cdots, 0.99 \}$ is given as well. This table can be compared to the Monte Carlo approximated quantiles obtained by \cite{durot03}.  All the code used in the numerical computations in this paper is available at \url{http://www.ceremade.dauphine.fr/~fadoua/bf2010_code/}.

\section{Monte Carlo approach}

We consider two different MC-based algorithms. They have the advantage of being very easy to understand and implement. %
The first approach is straightforwardly based on the expression of the distribution function of $M$ given in (\ref{ExpMonteCarlo}).  Because of the infinite product in (\ref{ExpMonteCarlo}), a first approximation due to the truncation of the product is introduced.  Control of the error due to this approximation is important in order to obtain a good theoretical estimator. Let $J > 0$ be some finite integer and consider the problem of estimating
$$
F_J(x) = E\left[\prod_{j=1}^J F_3 \bigg( \frac{x}{\sqrt{L_j}} \bigg) \right], \ \ \forall \ x > 0.
$$

For $x > 0$ and a given  $\epsilon \in (0, 1/4)$, the following lemma gives a lower bound for $J$ so that
\begin{eqnarray}\label{ApproxJ}
0 \le F_J(x) - F(x) < 2\epsilon.
\end{eqnarray}

\medskip

\begin{lemma}\label{LemmaApproxJ} The approximation error satisfies (\ref{ApproxJ}) if
\begin{eqnarray}\label{Jinf}
J \ge J_0 = \bigg  \lfloor \frac{-\log(x^2\epsilon^2/2)}{\log(2)} \bigg \rfloor + 1.
\end{eqnarray}

\end{lemma}

\medskip

\par \noindent \textbf{Proof.} See Appendix.

\medskip

For  $x > 0$ and a given $\epsilon > 0$, we draw $C$ independent copies $(L^{(c)}_1, L^{(c)}_2, L^{(c)}_3, \cdots, L^{(c)}_J)$ for $c=1, \cdots, C$ to estimate $F_J(x)$ where $J = J_0$ as given in Lemma \ref{ApproxJ}. The resulting Monte Carlo estimator is
\begin{eqnarray*}
\widehat F_{J,C}(x) = \frac{1}{C} \sum_{c=1}^C \prod_{j=1}^J F_3\bigg(\frac{x}{\sqrt{L^{(c)}_j}}\bigg).
\end{eqnarray*}
The computation of the distribution function $F_3$ imposes yet another approximation due to the fact that it is defined through an infinite series. The number of terms in the approximating finite sum needs to be larger for smaller values of $x$. Now by the Central Limit Theorem, we have
\begin{eqnarray*}
\sqrt{C} (\widehat F_{J,C}(x) - F_J(x))  \to_d \mathcal{N}(0, \sigma^2_J)
\end{eqnarray*}
with
$$
\sigma^2_J = Var\left[\prod_{j=1}^J F_3\bigg(\frac{x}{\sqrt{L_j}}\bigg)\right].
$$
Let $ z_{1-\alpha/2} $ be the $(1-\alpha/2)$- quantile of a standard normal for some small $\alpha \in (0,1)$. Then, for $C$ large enough the event
\begin{eqnarray*}
F_J(x)  \in \left[\widehat F_{J, C}(x) - \frac{\sigma_J \ z_{1-\alpha/2}}{\sqrt{C}}, \widehat F_{J, C}(x) + \frac{\sigma_J \  z_{1-\alpha/2}}{\sqrt{C}} \right]
\end{eqnarray*}
occurs with probability $ \approx 1-\alpha$. Combining both the deterministic and Monte Carlo approximations and noting that $\sigma^2_J \in [0,1]$, it follows that
\begin{eqnarray*}
F(x)  \in \left[\widehat F_{J, C}(x)  - 2\epsilon - \frac{z_{1-\alpha/2}}{\sqrt{C}}, \widehat F_{J, C}(x) + \frac{z_{1-\alpha/2}}{\sqrt{C}} \right]
\end{eqnarray*}
occurs with at least probability $\approx 1-\alpha$.  Hence, to ensure an error of order $\epsilon$, the sample size $C$ should be chosen of order $\lfloor 1/\epsilon^2 \rfloor$. Therefore, very large sample sizes are needed to get accurate results. To give an order of magnitude, Table \ref{J0C} shows several values of $J_0$ and $C$ corresponding to desired precision targets. All the values are computed for $ 0.33 \le x \le 2.54 $, where 0.33 appears to be the numerical limit of what we can compute without violating the basic properties of a distribution function. This point will be brought up again in the next section. Note that the main purpose of Table \ref{J0C} is to give an idea about how $J_0$ and $C$ behave as functions of the precision. For instance,  a precision of order $10^{-5}$ is useless if the goal is to compute an approximation of the value distribution function of $M$ at $0.33$ since it is of order $10^{-12}$ as found with the GS algorithm.

We use the above MC approach to estimate the distribution function of $M$ for $0.33 \le x \le 2.54$ as well as the upper quantiles. The algorithm is implemented in C. This method turns out to be very slow for large sample sizes.  Moderate sample sizes (of order $10^6$) do not give the desired accuracy for small $x$. The estimates of the distribution function for large $x$ (of order 0.80 and above) as well as the upper quantiles match with those obtained by GS algorithm (see next section).

\bigskip

In the same vein, one can consider a second variant of MC. It is mainly based on the following result due to Kennedy 1976 (see Corollary on page 372):
$$
M_3 =_d \sup_{t \in [0,1]} \br(t) - \inf_{t \in [0,1]} \br(t)
$$
where $\br$ is a Brownian bridge on length 1. Now using the well-known Donsker approximation, the distribution of $M_3$ can be approximated for large $N$ by the distribution of the random variable
$$
V_N =  \sup_{t \in [0,1]} \sqrt{N}  (\mathbb G_N(t) - t)  - \inf_{t \in [0,1]} \sqrt{N}  (\mathbb G_N(t) - t)
$$
where $\mathbb G_N$ is the uniform empirical process based on $N$ independent uniform random variables $U_1, \cdots, U_N$ in $[0,1]$. Using the fact that $\mathbb G_N $ is a constant function between the order statistics $U_{(1)} < \cdots < U_{(n)}$, it can be easily shown that
$$
V_N = \sqrt{N} \left \{ \max_{1 \le i \le N} \left(\frac{i}{N} - U_{(i)}\right) - \min_{1 \le i \le N } \left(\frac{i-1}{N} - U_{(i)}\right) \right\}.
$$
Now the formula in (\ref{MainTheo}) yields the weak approximation
$$
M_{J, N} = \max_{1 \le i \le J} \sqrt{L}_j V^{(j)}_N
$$
where $V^{(1)}_N, V^{(2)}_N, \cdots$ are independent random variables distributed as $V_N$, and $J$ is a positive integer that should be chosen large enough to have the truncation error under control as done above. The distribution function of $M$ can be estimated empirically by generating $C$ independent random variables $M^{(1)}_{J, N}, M^{(2)}_{J, N}, \cdots,  M^{(C)}_{J, N}$ with the same distribution as $M_{J, N}$.
If this second variant of MC has the drawback of adding another error due to the stochastic approximation of $F_3$ by that of $V_N$, it gives the possibility to generate samples with a distribution close to that of $M$ for $J$, $N$ and $C$ large enough. We will not pursue  here the calculation of the approximation error as a function of $J$, $N$ and $C$, which have to be very large to achieve high precision.

\begin{figure*}
\begin{center}
\includegraphics[width=0.65\textwidth]{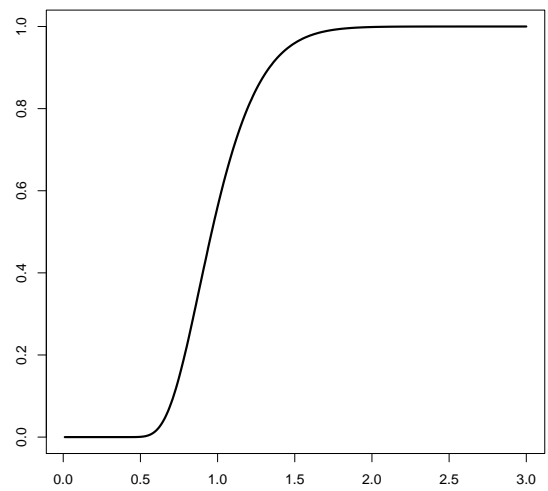}
\caption{Plot of a Monte Carlo approximation of $F$ based on a sample of size 10,000, with $J_0= 100$.}
\label{MCF}
\end{center}
\end{figure*}

\begin{figure*}
\begin{center}
 \includegraphics[width=0.65\textwidth]{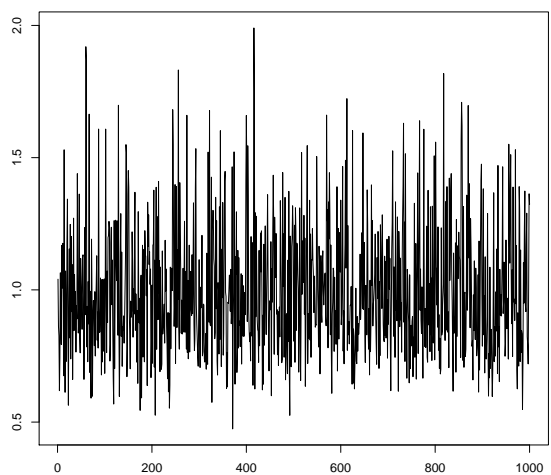}
\caption{Plot of the trajectory of random sample of size 1000 of independent realizations of $M_{J, N}$ with $J=100$ and $N=10,000$.}
\end{center}
\label{Simu1000}       %
\end{figure*}

The plot in Figure \ref{MCF} shows an estimation of $F$ using the first MC method with $J_0= 100$ and $C=10,000$. If the values are not accurate for small $x$, the plot gives nevertheless a good idea about the true shape of $F$. This is confirmed by the approximation results we obtain with the numerical inversion of Laplace transform. The trajectory of 1000 independent random variables with the same distribution of $M_{J,N}$ for $J=100$ an $N = 10,000$ is shown in Figure \ref{Simu1000}. The sample was extracted from a larger one of size 10,000 with an empirical mean and standard deviation equal to $ 0.9970$ and  $0.2475$ respectively.

If the MC approach gives a first idea of the support and shape of the distribution of $M$, it is not satisfactory in terms of efficiency and precision. As we show in the next section, the GS algorithm is a much better choice in both respects.

\section{Gaver-Stehfest algorithm}

The Gaver-Stehfest (GS) algorithm is one of several algorithms of numerical inversion of Laplace transform. For an excellent description of these algorithms, see \cite{Abatwhitt06}. The GS algorithm is different from other inversion procedures in that it involves only real numbers, but it also requires a very high numerical precision as we explain below (also see~\cite{Abatwhitt06}, p. 415).
If $g$ is the Laplace transform of some function $f$ defined on $\mathbb R$, then GS approximation of $f$ is given by
\begin{eqnarray}\label{ApproxInvLT}
\tilde{f}_K(t) = \frac{\ln(2)}{t} \sum_{k=1}^{2K} \xi_k g\left( \frac{k \ln(2)}{t} \right)
\end{eqnarray}
where $K$ is an integer in $\mathbb N^*$ and
\begin{eqnarray*}
 \xi_k &= & \frac{(-1)^{k+K}}{K!} \sum_{j=\lfloor (k+1)/2 \rfloor }^{k \wedge K} j^{K+1} {K\choose j} {2j \choose j} {j \choose k-j}, \ \ 1 \le k \le 2K.
\end{eqnarray*}
Under the assumption that the inverse of Laplace transform $f$ has all its singularity points in $(-\infty, 0]$ and that is infinitely differentiable on $(0, \infty)$, an extensive computation study carried out by \cite{abatevalko04} has shown that
\begin{eqnarray*}
\left \vert \frac{\tilde{f}_K(t) - f(t)}{f(t)} \right \vert  \approx 10^{-0.8K}, \ \ t > 0.
\end{eqnarray*}
If the function $f$ is bounded by 1 say, then the approximation in (\ref{ApproxInvLT}) for well-behaved functions (in the sense given above) coincides with the truth up to significant $0.8K$ digits. Hence, the bigger $K$ is, the better is the approximation. However, for large values of $K$, the binomial coefficients in $\xi_k$ become extremely large and require high numerical precision.  Such a facility is typically provided by a Multiple Precision (MP) numerical library or is built-in in some programming languages.

For a given integer $K > 0$, let $\tilde F_{K}$ denote the GS approximation of $F$.  From the formula of $F$ in (\ref{ExpInvLT}) and (\ref{ApproxInvLT}), it is easily seen that
\begin{eqnarray}\label{FK}
\tilde F_{K}(x) = \sum_{k =1}^{2K} \frac{\xi_k}{k} \ G(\sqrt{k \log(2)} \ x), \ \ \ x > 0
\end{eqnarray}
where $G$ is the same function defined by the infinite product in (\ref{G}).

\bigskip

For $x > 0$ and a given $\epsilon > 0$ we approximate $G$ by the product of the first $N$ terms, where $N$ is a positive integer depending on $x$ and $\epsilon$. Define
$$
G_N(t) = \prod_{j=1}^N \exp\left( -4 \left( 2 \sqrt{2} t n K_1(2 \sqrt{2} n t) - K_0(2 \sqrt{2}n t) \right) \right), \ \ t > 0
$$
the truncated version of $G$.  This truncation induces an additional error which we need to control. In fact, in computing the Gaver-Stehfest approximation of the distribution function $F$, we actually replace $\tilde{F}_K$ in (\ref{FK}) by
\begin{eqnarray}\label{FNK}
\tilde F_{N, K}(x) = \sum_{k =1}^{2K} \frac{\xi_k}{k} \ G_N(\sqrt{k \log(2)} \ x), \ \ \ x > 0.
\end{eqnarray}
The following shows that the error due to replacing $\tilde{F}_K(x)$ by $\tilde{F}_{N, K}(x)$ does not exceed a given threshold $\epsilon > 0$ provided than $N$ is large enough.

\medskip

\begin{lemma}\label{ApproxG}
For $\epsilon > 0$, we have $\vert \tilde{F}_{N, K}(x) - \tilde{F}_K(x) \vert \le \epsilon$ if $N \ge N_0$ where
\begin{eqnarray}\label{LowerN}
N_0 & = & \left \lfloor \frac{1}{\sqrt{2 \ln(2)} \ x } \left\{ \ln\left(\frac{1}{\epsilon (1 - \exp(- \sqrt{2 \ln(2)} \ x ))}\right) + (2K+1)\ln(K) + 3K+2 \right \} \right \rfloor \nonumber \\
&& + 1.   \nonumber \\
&&
\end{eqnarray}
\end{lemma}

\medskip

\par \noindent \textbf{Proof.} \ See Appendix.

\bigskip

From Lemma \ref{ApproxG} it follows that
\begin{eqnarray*}
\left \vert \tilde{F}_{N, K}(x) - F(x) \right \vert \le  \epsilon + \left \vert \tilde{F}_{K}(x) - F(x) \right \vert.
\end{eqnarray*}
The second term in the left side is known to be of order $10^{-0.8K}$, and hence the approximation is of the same order  if $\epsilon$ is chosen to be $o(10^{-0.8K})$, and of order $\epsilon$ if the latter dominates and $N$ is chosen to be larger or equal than $N_0$ given in (\ref{LowerN}).

\bigskip

\bigskip

We implement  the multiple precision  calculation of $\tilde{F}_{K}$
in  C++  using  two  open-source  libraries  for  arbitrary  precision
computation: the GNU Multiple Precision Arithmetic Library (see \cite{GMP})
and  the  Multiple Precision  Floating-point  Reliable Library  (MPFR); see
\cite{fousse07}. GMP is  an  optimized library  written  in C  with
assembly code for common inner loops.  MPFR is built on top of GMP and
adds support for common floating-point operations such as $\exp(x)$.

To  approximate  the Bessel  functions  in  (\ref{G}),  we use  Bessel
routines  from the  ALGLIB  library\footnote{Available    at
\url{http://www.alglib.net/specialfunctions/bessel.php}} based  on  piecewise rational  and
Chebyshev  polynomial  approximations. We  use  a
precision of 4000 bits  to represent multiple precision floating-point
numbers.   However,  the  provided  AGLIB Bessel  approximations  only
guaranty a maximal error  of order $10^{-14}$.  As a proof-of-concept,
we have  also implemented the same  algorithm using a  much slower but
more  accurate  numerical   library  in  Python\footnote{Available  at
\url{http://code.google.com/p/mathpy/}}.  For small  values of $x$ such as
0.30,  0.31, and  0.32,  and unlike  with  the C library,  we obtain  results
consistent with the monotonicity and positivity of a cumulative distribution function. For $K=60$, $N=3200$, the Python code gives the following approximations $9.8605317729e{-14}$ for $x=0.32$,  $1.10482969e{-12}$ for $x=0.32$ and $9.67030359e{-12}$ for $x=0.33$.

Computing $\tilde{F}_{K}(x), x \in [0.33, 2.54]$ takes about 6 hours
(90  seconds  per  function  evaluation) on  a  2GHz  single-processor
machine.   The  computation is  dominated  by  the  evaluation of  $G$
in (\ref{G}). The  coefficients $\xi_k, k=1, \cdots, 2K$ need to be computed only once. Tables \ref{DisFun1}, \ref{DisFun2} and \ref{DisFun3} give the approximated values of $F$ on a grid starting at 0.33 and ending at 2.54 with a regular mesh chosen to be equal 0.01.

Finally, computing the upper quantiles of order is crucial when using the Kolomogorov type monotonicity test based on the maximal distance between the empirical cumulative sum diagram (resp. the empirical distribution) in the regression estimation setting (resp. the density estimation setting), see \cite{durot03} and \cite{kuliklop08}. The GS algorithm can be easily used to approximate the upper quantiles of order $p \in \{ 0.90, 0.91, \cdots, 0.99 \}$.

Note  that  these quantiles  are  between  1.33  and 1.72  (see  Table
\ref{DisFun2}).  For each quantile, we used a binary search and stopped when
the difference  between the GS approximation  of $F$ at  the point and
the targeted  probability falls below a given  threshold ($10^{-7}$ in
the results we report).
The results are  shown in Table \ref{Quantiles}.  This  table is to be
compared  with the  one published by \cite{durot03} who  obtained the
quantiles for the same probabilities using a Monte Carlo approach.

\bigskip

In this paper, Monte Carlo and a numerical inversion of the Laplace transform were used to estimate the distribution function and upper quantiles of $M$, the maximal difference between a Brownian motion on $[0,1]$ (or a Brownian bridge of length 1) and its concave majorant. This random variable determines the asymptotic critical region of a nonparametric test for monotonicity of a density or regression curve. We find the numerical inversion of Laplace transform,  based here on the Gaver-Stehfest algorithm, to be much more accurate and faster than the Monte Carlo method. Numerical inversion of Laplace transform was then very well adapted to this problem. However, it would not have been possible to use such an efficient method if a Laplace transform representation of the distribution of $M$ was not available, see \cite{balabdpitman09}.

Finally, we would like to draw the reader's attention to the earlier computational work of \cite{groewell01} on Chernoff's distribution. The latter appears as the limit distribution of the Grenander estimator; that is the Maximum Likelihood estimator of a decreasing density on $(0, \infty)$. In their work, \cite{groewell01} have also used a mathematical characterization of Chernoff's distribution. This allowed for a very efficient and fast approximation procedure which also outperformed Monte Carlo estimation.

\begin{figure*}
\begin{center}
\includegraphics[width=1\textwidth]{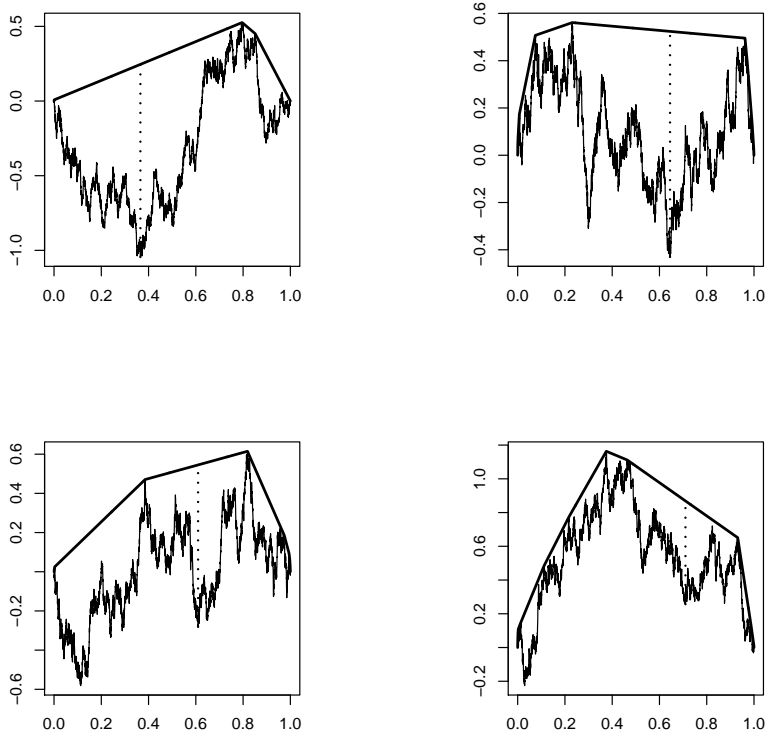}
\caption{Four realizations of Brownian bridge and its concave majorant.  The length of the dotted vertical segment equals $M$, the realization of the maximum difference between the Brownian bridge and its concave majorant. The Haar approximation was used to generate the Brownian bridge on a discrete partition of $[0,1]$ with a mesh equal to $2^{-12}$.}
\label{fig: 4BrandCM}
\end{center}
\end{figure*}

\begin{table}[!h]
\caption{Order of the lower bound $J_0$ and sample size $C$.}
\label{J0C}
\begin{center}
\begin{tabular}{c|c|c}
\hline
\hline
Precision  & $J_0$ &  $C$ \\
$10^{-5}$ &  38 & $10^{-10}$ \\
$10^{-8}$ &  57 &  $10^{-16}$  \\
$10^{-10} $  & 71 & $10^{-20}$ \\
$10^{-20} $  & 138 & $10^{-40} $ \\
\hline
\hline
\end{tabular}
\end{center}
\end{table}

\begin{table}[htbp]
\caption{Approximated values of $F$ obtained by the Gaver-Stehfest algorithm with $K =100$, $N=3200$.} %
\label{DisFun1}
\begin{center}
\begin{tabular}{cc|cc|cc}
\hline
\hline
$x$ & $\tilde{F}_{K}$ &  $x$ & $\tilde{F}_{K}$ &  $x$ & $\tilde{F}_{K}$ \\
\hline
0.33	& 9.24257424322e-12 &  0.61 &	1.90415418636e-2 & 0.89 &	3.75627377529e-1 \\
0.34  & 	3.74465832649e-10 &  0.62 &	2.34012345624e-2  & 0.90 & 	3.93227821925e-1 \\
0.35	&   1.96857226601e-9 &  0.63 & 	2.84117820291e-2  & 0.91 &	4.10762393335e-1 \\
0.36 & 	 8.39648450672e-9  &  0.64 & 	3.41073192977e-2  & 0.92 &	4.28198510241e-1 \\
0.37	& 3.13734889037e-8 &  0.65 & 	4.05156243499e-2  & 0.93 &	4.45505815829e-1 \\
0.38  &	1.04444192578e-7 & 0.66  &	4.76576519986e-2 & 0.94 &	4.62656184022e-1  \\
0.39 &	3.13540455103e-7 &  0.67 & 	5.55472601039e-2 & 0.95 &	4.79623704507e-1 \\
0.40 & 	8.57933368778e-7 &  0.68 &	6.41911192791e-2 & 0.96 & 	4.96384649825e-1 \\
0.41 & 	2.16022136991e-6 &  0.69 &	7.35887911016e-2 & 0.97 & 	5.12917427370e-1  \\
0.42 & 	5.04741838646e-6  &  0.70 & 8.37329554403e-2 & 0.98 &	5.29202518946e-1 \\
0.43 & 	1.10248951989e-5 & 0.71	& 9.46097647495e-2  & 0.99 & 	5.45222410267e-1  \\
0.44 & 	2.26592425775e-5 & 0.72 & 	1.06199301858e-1 & 1.00 &	5.60961512572e-1 \\
0.45 & 	4.40745827182e-5 &  0.73 & 	1.18476117686e-1 &  1.01 &	5.76406078288e-1  \\
0.46 & 	8.15505378035e-5 &  0.74 &	1.31409826206e-1 & 1.02 & 	5.91544112452e-1  \\
0.47 & 	1.44191509163e-4  & 0.75 &	1.44965735553e-1 & 1.03 & 	6.06365281379e-1 \\
0.48 & 	2.44619527193e-4  & 0.76 &	1.59105496305e-1 & 1.04 & 	6.20860819881e-1  \\
0.49 & 	3.99630103610e-4 &  0.77 &	1.73787750362e-1 & 1.05 & 	6.35023438140e-1  \\
0.50 & 	6.30744831947e-4 &  0.78 &	1.88968766384e-1 & 1.06 & 	6.48847229167e-1  \\
0.51 & 	9.64597141746e-4 &  0.79 & 	2.04603050312e-1 & 1.07 & 	6.62327577632e-1  \\
0.52 & 	1.43309828337e-3 & 0.80 &	2.20643921928e-1 & 1.08 &	6.75461070708e-1  \\
0.53 &	2.07334764606e-3 & 0.81 & 	2.37044050697e-1 & 1.09 & 	6.88245411417e-1 \\
0.54 & 	2.92727237058e-3 &  0.82 &	2.53755946188e-1 &1.10 & 	7.00679334912e-1  \\
0.55 &	4.04100308041e-3 &  0.83 & 	2.70732400181e-1 &  1.11 &	7.12762527962e-1\\
0.56 &	5.46401279238e-3 &  0.84 & 	2.87926879124e-1 & 1.12 &	7.24495551883e-1  \\
0.57 & 	7.24806261908e-3 &  0.85 & 	3.05293866924e-1 & 1.13 &	7.35879769041e-1 \\
0.58 & 	9.44600944518e-3 &  0.86 &	3.22789159067e-1 & 1.14 &	7.46917273011e-1  \\
0.59 &	1.21105368429e-2 &  0.87 & 	3.40370109977e-1& 1.15  &	7.57610822418e-1 \\
0.60 & 	1.52928712783e-2 &  0.88  & 3.57995836076e-1 & 1.16 & 	7.67963778457e-1 \\
\hline
\end{tabular}
\end{center}
\end{table}

\begin{table}[htbp]
\caption{Approximated  values of $F$ obtained by the Gaver-Stehfest algorithm.}
\label{DisFun2}
\begin{center}
\begin{tabular}{cc|cc|cc}
\hline
\hline
$x$ & $\tilde{F}_{K}$ &  $x$ & $\tilde{F}_{K}$ &  $x$ & $\tilde{F}_{K}$ \\
\hline
1.17	& 7.77980046011e-1 & 1.51 &	9.62020662224e-1 & 1.85 &	9.96016083423e-1 \\
1.18	& 7.87664018322e-1 & 1.52 &	9.64212812296e-1 & 1.86 &	9.96298535959e-1 \\
1.19	& 7.97020525083e-1 & 1.53 &	9.66292598240e-1 & 1.87 &	9.96562371771e-1 \\
1.20	& 8.06054783852e-1 & 1.54 &	9.68264838078e-1 & 1.88 &	9.96808708806e-1 \\
1.21	& 8.14772354647e-1 & 1.55 & 9.70134203687e-1	 & 1.89 & 9.97038605986e-1	 \\
1.22	& 8.23179097587e-1 & 1.56 &	9.71905221291e-1 & 1.90 &	9.97253065744e-1 \\
1.23	& 8.31281133430e-1 & 1.57 &	9.73582272276e-1 & 1.91 &	9.97453036494e-1 \\
1.24	& 8.39084806872e-1 & 1.58 &	9.75169594296e-1 & 1.92 &	9.97639415022e-1 \\
1.25	& 8.46596652448e-1 & 1.59 &	9.76671282637e-1 & 1.93 &	9.97813048825e-1 \\
1.26	& 8.53823362907e-1 & 1.60 &	9.78091291833e-1 & 1.94 &	9.97974738360e-1 \\
1.27	& 8.60771759907e-1 & 1.61 &	9.79433437494e-1 & 1.95 &	9.98125239231e-1 \\
1.28	& 8.67448766898e-1 & 1.62 &	 9.80701398339e-1& 1.96 &	9.98265264308e-1 \\
1.29	& 8.73861384073e-1 &  1.63 & 9.81898718407e-1 & 1.97 &9.98395485767e-1	 \\
1.30	& 8.80016665251e-1 & 1.64 &	9.83028809430e-1 & 1.98 &	9.98516537064e-1 \\
1.31	& 8.85921696569e-1 & 1.65 &	9.84094953345e-1 & 1.99 &	9.98629014836e-1 \\
1.32	& 8.91583576893e-1 & 1.66 &	9.85100304937e-1 & 2.00 &	9.98733480735e-1 \\
1.33	& 8.97009399815e-1 & 1.67 &	9.86047894590e-1 & 2.01 &	9.98830463190e-1 \\
1.34	& 9.02206237159e-1 & 1.68 &	9.86940631128e-1 & 2.02 &	9.98920459107e-1 \\
1.35	& 9.07181123890e-1 &  1.69&	9.87781304739e-1 & 2.03 &	9.99003935491e-1 \\
1.36	& 9.11941044337e-1 & 1.70 &	9.88572589969e-1 & 2.04 &	9.99081331015e-1 \\
1.37	& 9.16492919660e-1 & 1.71 &	9.89317048762e-1 & 2.05 &	9.99153057516e-1 \\
1.38	& 9.20843596472e-1 & 1.72 &	9.90017133547e-1 & 2.06 &	9.99219501431e-1 \\
1.39	& 9.24999836546e-1 & 1.73 &	9.90675190351e-1 & 2.07 &	9.99281025174e-1 \\
1.40	& 9.28968307546e-1 & 1.74 &	9.91293461936e-1 & 2.08 &	9.99337968446e-1 \\
1.41	& 9.32755574715e-1 &  1.75&	9.91874090944e-1 & 2.09 &	9.99390649494e-1 \\
1.42	& 9.36368093452e-1 & 1.76 &	9.92419123041e-1 & 2.10 &	 9.99439366308e-1 \\
1.43	& 9.39812202742e-1 &  1.77 & 9.92930510053e-1	 & 2.11 &	9.99484397768e-1 \\
1.44	& 9.43094119365e-1 & 1.78 &	9.93410113095e-1 & 2.12 &	9.99526004728e-1 \\
1.45	& 9.46219932852e-1 & 1.79 &	9.93859705663e-1 & 2.13 &	9.99564431063e-1 \\
1.46	& 9.49195601129e-1 &  1.80&	9.94280976712e-1 & 2.14 & 9.99599904647e-1	 \\
1.47	& 9.52026946811e-1 & 1.81 &	9.94675533688e-1 & 2.15 &	9.99632638303e-1 \\
1.48	& 9.54719654107e-1 & 1.82 &	9.95044905522e-1 & 2.16 & 9.99662830687e-1	 \\
1.49	& 9.57279266289e-1 & 1.83 &	9.95390545586e-1 & 2.17 &	9.99690667143e-1 \\
1.50	& 9.59711183695e-1 &  1.84 & 9.95713834588e-1	 & 2.18 &	9.99716320502e-1 \\
\hline
\end{tabular}
\end{center}
\end{table}

\begin{table}[htbp]
\caption{Approximated values of $F$ obtained by the Gaver-Stehfest algorithm.}
\label{DisFun3}
\begin{center}
\begin{tabular}{cc}
\hline
\hline
$x$  &  $\tilde{F}_{K}(x)$  \\
\hline
2.19 &	9.99739951848e-1 \\
2.20 &	9.99761711238e-1 \\
2.21 &	9.99781738392e-1 \\
2.22 &	9.99800163331e-1 \\
2.23 &	9.99817106996e-1 \\
2.24 &	9.99832681825e-1 \\
2.25 &	9.99846992298e-1 \\
2.26 &	9.99860135450e-1 \\
2.27 &	9.99872201360e-1 \\
2.28 &	9.99883273608e-1 \\
2.29 &	9.99893429703e-1 \\
2.30 &	9.99902741490e-1 \\
2.31 &	9.99911275534e-1 \\
2.32 &	9.99919093474e-1 \\
2.33 &	9.99926252361e-1 \\
2.34 &	9.99932804973e-1 \\
2.35 &	9.99938800114e-1 \\
2.36 &	9.99944282889e-1 \\
2.37 &	9.99949294965e-1 \\
2.38 &	9.99953874813e-1 \\
2.39 &	9.99958057939e-1 \\
2.40 &	9.99961877094e-1 \\
2.41 &	9.99965362474e-1 \\
2.42 &	9.99968541907e-1 \\
2.43 & 	9.99971441026e-1 \\
2.44 &	9.99974083431e-1 \\
2.45 &	9.99976490838e-1 \\
2.46 &	9.99978683223e-1 \\
2.47 &	9.99980678951e-1 \\
2.48 &	9.99982494897e-1 \\
2.49 &	9.99984146562e-1 \\
2.50 &	9.99985648176e-1 \\
2.51 & 	9.99987012798e-1 \\
2.52 &	9.99988252405e-1 \\
2.53 &	9.99989377977e-1 \\
2.54 &	9.99990399575e-1 \\
\hline
\end{tabular}
\end{center}
\end{table}

\begin{table}[!h]
\caption{Approximated upper quantiles $q_{1-\alpha}$ of order $1 - \alpha$. The approximation is based on the Gaver-Stehfest algorithm with $K =100$, $N=60$.} %
\label{Quantiles}
\begin{center}
\begin{tabular}{cccccc}
\hline
\hline
$\alpha$ & 0.01 &  0.02 & 0.03 & 0.04 & 0.05  \\
$q_{1-\alpha}$ & 1.71974853  &  1.61439819 &  1.54926391 & 1.50122253 & 1.46279052  \\
\hline
 &  0.06 &  0.07 & 0.08 & 0.09 & 0.10  \\
 & 1.43055908 & 1.40267791 & 1.37802490 & 1.35586822 & 1.33570159 \\
\hline
\hline
\end{tabular}
\end{center}
\end{table}

\newpage

\section*{Appendix}

The following facts will be used in the proof of Lemma \ref{LemmaApproxJ}.

\medskip

\par \noindent \textbf{Lemma A.1} \ \textit{We have
\begin{description}
\item [(i)] For all $j \in \mathbb N^*$, $E(L_j) = 1/2^{j}$.
\item [(ii)] For $x \ge \sqrt{2}$, $F_3(x) \ge \exp(-1/x^2)$.
\end{description}
}

\medskip

\paragraph{Proof.} \ The first identity can be proved recursively. For $j=1$, we have $E(L_1) = E(U_1) =1/2$. Suppose that $E(L_i) = 1/2^i$ for all $i \le j$. It is easy to check that
$$
L_{j+1} = (1-U_1)(1-U_2)\cdots (1-U_j) U_{j+1} = (1- \sum_{i=1}^j L_i) U_{j+1}.
$$
By independence of $(L_1, \cdots, L_j)$ and $U_{j+1}$, we can write
$$
E(L_{j+1}) = E(1 - \sum_{i=1}^j L_i) /2 = (1 - \sum_{i=1}^j 1/2^i)/2 = 1/2^{j+1}
$$
and the identity is proved for all $j \in \mathbb N^*$.

For the second inequality, we will use the fact that for a given $a \ge 1$
\begin{eqnarray}\label{IneqInterm}
(4a^2 t - 1) \exp(-2a^2 t) \le \exp(- a t), \ \ \textrm{for all $t \ge 0$}.
\end{eqnarray}
Consider the function
\begin{eqnarray*}
h(t): = (4a^2t - 1) \exp(-a(2a-1)t), \ \ t \ge  0.
\end{eqnarray*}
The study of variations of $h$ shows that $h$ is increasing on $[0, (6a-1)/(2a-1)]$ and decreasing on $[(6a-1)/(2a-1), \infty)$ with with $h(0) = - 1$,  $\lim_{t \to \infty} h(t) =0$ and
$$
h\left(\frac{6a-1}{2a-1}\right) = \frac{4a}{2a-1} \exp\left(-\frac{6a-1}{4a}\right).
$$
Now, the function
\begin{eqnarray*}
\log h\left(\frac{6a-1}{2a-1}\right) = \log\left( \frac{4a}{2a-1} \right)  - \frac{6a-1}{4a}
\end{eqnarray*}
is decreasing on $[1,  \infty)$ with $\log h(1) = \log(4) - 5/4 < 0$, and hence $h((6a-1)/(2a-1)) < 1$. It follows that $h(t) < 1$ and the inequality in (\ref{IneqInterm}) is proved.

It follows that
\begin{eqnarray*}
1-F_3(x) = 2 \sum_{k=1}^\infty (4k^2 x^2 - 1) \exp(-2k^2 x^2) & \le &  2 \sum_{k=1}^\infty \exp(- kx^2) \\
                                                               & = &  \frac{2\exp(-x^2)}{1 -\exp(-x^2)}.
\end{eqnarray*}
To show that $F_3(x) \ge \exp(-1/x^2)$ for all $x \ge 2$, it is enough to show that
\begin{eqnarray*}
 \frac{1 - 3\exp(-x^2)}{1 -\exp(-x^2)} \ge \exp(-1/x^2), \ \ \textrm{for all $x\ge \sqrt{2}$}
\end{eqnarray*}
or equivalently
\begin{eqnarray*}
2 \exp(-t) \le (1-\exp(-t))(1-\exp(-1/t)), \ \ \textrm{for all $t\ge 2$}.
\end{eqnarray*}
The preceding inequality can be proved as follows. Define the function
\begin{eqnarray*}
k(t):= (\exp(t) - 1)(1-\exp(-1/t)) , \ \ t \ge 2.
\end{eqnarray*}
We will show now that $k(t) \ge 2$ for all $t \ge 2$.  For $t \ge 2$, we have
\begin{eqnarray*}
k'(t)& = &  \exp(t)\left \{1 - \left(1 + \frac{1}{t^2} \right) \exp(-1/t) \right\} + \frac{\exp(-1/t)}{t^2} \\
     & \ge & \exp(t)\left\{1 - \left(1 + \frac{1}{t^2} \right) \exp(-1/t) \right\} \\
     & = &  \exp(t) \phi(1/t)
\end{eqnarray*}
where
\begin{eqnarray*}
\phi(z) = 1 - (1 + z^2) \exp(-z), \ \ z \in [0, 1/2].
\end{eqnarray*}
It is easy to show that $\phi$ is increasing on $[0,1/2]$ and hence $\phi(z) \ge \phi(0) = 1$. It follows that the function $k$ is increasing on $[2, \infty)$. Since $k(2) \approx  2.514 \ge 0$, the inequality $F_3(x) \ge \exp(-1/x^2), \ x \ge \sqrt{2}$ follows.   \hfill  $\Box$

\paragraph{Proof of Lemma \ref{LemmaApproxJ}.} \  Define
$$
\Delta_J := 1 -  \prod_{j=J+1}^\infty F_3 \bigg( \frac{x}{\sqrt{L_j}} \bigg).
$$
We have
\begin{eqnarray*}
0 \le F_J(x) - F(x) & = & E\left[\prod_{j=1}^J F_3 \bigg( \frac{x}{\sqrt{L_j}} \bigg) \Delta_J  \right] \\
                    & \le & E[\Delta_J] \\
                    &  =  &  E\left[\Delta_J 1_{\Delta_J \le \epsilon}\right] +   E\left[\Delta_J 1_{\Delta_J > \epsilon}\right] \\
                    & \le & \epsilon + P(\Delta_J > \epsilon).
\end{eqnarray*}
Let $A_J$ be the event
$$
A_J = \left \{ L_j \le x^2/4, \ \textrm{for all $j \ge J+1$} \right \}
$$
and $A^c_J$ its complement.

We can write
\begin{eqnarray*}
P(\Delta_J > \epsilon) &= & P\bigg(\prod_{j = J+1}^\infty F_3\bigg( \frac{x}{\sqrt{L_j}} \bigg) < 1 - \epsilon \bigg) \\
                       & = &  P\bigg(\bigg \{\prod_{j = J+1}^\infty F_3\bigg( \frac{x}{\sqrt{L_j}} \bigg) < 1 - \epsilon  \bigg \} \cap A_J\bigg) + P(A^c_J) \\
                       & \le & P\bigg(\prod_{j = J+1}^\infty \exp(-L_j/x^2) < 1 - \epsilon \bigg) +  P(A^c_J), \ \textrm{using Lemma A.1 (i)} \\
                       & = & P\bigg( \sum_{j = J+1}^\infty L_j >  x^2 \log(1/(1 - \epsilon)) \bigg) +  P(A^c_J) \\
                       & \le & P\bigg( \sum_{j = J+1}^\infty L_j >  x^2 \log(1/(1 - \epsilon)) \bigg) +  \sum_{j=J+1}^\infty P(L_j > x^2/4).
                      \end{eqnarray*}
Using Lemma A.1 (ii) and the Chebyshev inequality, we get
\begin{eqnarray*}
P\bigg( \sum_{j = J+1}^\infty L_j >  x^2 \log(1/(1 - \epsilon)) \bigg) &  \le &  \frac{\sum_{j=J+1}^\infty 1/2^j}{x^2 \log(1/(1 - \epsilon))} \\
                                                                       & = &  \frac{1}{2^{J} x^2 \log(1/(1 - \epsilon))}
\end{eqnarray*}
and
\begin{eqnarray*}
\sum_{j=J+1}^\infty P(L_j > x^2/4) \le  \frac{4}{2^J x^2}.
\end{eqnarray*}
Hence,
\begin{eqnarray*}
0 \le F_J(x) - F(x) & \le & \epsilon + \frac{1}{2^{J} x^2 \log(1/(1 - \epsilon))} + \frac{4}{2^J x^2}.
\end{eqnarray*}
To have this approximation error smaller than $ 2 \epsilon$, it suffices to take
\begin{eqnarray*}
J > \frac{1}{\log(2)} \left( \log\bigg(\frac{1}{\epsilon x^2}\bigg) + \log\bigg( \frac{1}{\log(1/(1-\epsilon))} + 4 \bigg)  \right).
\end{eqnarray*}
If $\epsilon < 1/4$ we can take
\begin{eqnarray}\label{Jinf}
J \ge \bigg \lfloor \frac{-\log(x^2\epsilon^2/2)}{\log(2)} \bigg \rfloor + 1
\end{eqnarray}
and Lemma \ref{ApproxJ} is proved.  \hfill $\Box$

\paragraph{Proof of Lemma \ref{ApproxG}. } \ The modified Bessel function of the second kind $K_n$ is known to converge to 0 as $x \to \infty$. Moreover we have
\begin{eqnarray*}
K_n(x) = \sqrt{\frac{\pi}{2}} \left(\frac{\exp(-x)}{\sqrt{x}} + o\left(\frac{1}{x}\right)\right), \ \ x > 0,
\end{eqnarray*}
and \begin{eqnarray*}
xK_1(x) - K_0(x)  \le \sqrt{\frac{\pi}{2}} \sqrt{x} \exp(-x),  \ \ \forall x > 0
\end{eqnarray*}
see Lemma A.2. For $ t > 0$, define
\begin{eqnarray*}
H(t) = 4 \sum_{n=1}^\infty \left(2\sqrt{2} n t K_1(2\sqrt{2} n t) - K_0(2\sqrt{2} n t) \right)
\end{eqnarray*}
so that $G(t) = \exp(-H(t))$. Also, for $N \in \mathbb N^*$ let
\begin{eqnarray*}
H_N(t) = 4 \sum_{n=1}^N \left(2\sqrt{2} n t K_1(2\sqrt{2} n t) - K_0(2\sqrt{2} n t) \right)
\end{eqnarray*}
so that $G_N(t) = \exp(-H_N(t))$. We have
\begin{eqnarray}\label{FondIneq}
0 < G_N(t) - G(t) = \exp(-H_N(t)) - \exp(-H(x)) &\le & H(t) - H_N(t) \nonumber \\
                                                 & = & 4 \sum_{n=N}^\infty \left(nc K_1(nc) - K_0(nc)\right) \nonumber \\
                                                 & \le &  4 \sqrt{\frac{\pi}{2}} \frac{\exp(-c N/2)}{1 - \exp(-c/2)}
\end{eqnarray}
where $c = 2 \sqrt{2} t$.

Let us write again $\tilde{F}_{N, K}$ for the Gaver-Stehfest approximation of the inverse of Laplace transform of $G_N$.  %

The corresponding approximation error due to truncating $G$ is given by
\begin{eqnarray*}
\tilde{E}_{N, K}(x) = \tilde{F}_K(x) - \tilde{F}_{N, K}(x) = \sum_{k=1}^{2K} \frac{\xi_k}{k} \left(G(\sqrt{k \ln(2)} \ x)  - G_N(\sqrt{k \ln(2)} x) \right), \ \ x > 0.
\end{eqnarray*}
By (\ref{FondIneq}), we can write
\begin{eqnarray*}
\left \vert \tilde{E}_{N, K}(x) \right \vert = 4 \sqrt{\frac{\pi}{2}}\sum_{k=1}^{2K} \frac{\vert \xi_k \vert}{k} \frac{\exp(-\alpha_k N)}{1 - \exp(-\alpha_ k)}
\end{eqnarray*}
where
$\alpha_k = \sqrt{2 \ln(2) k} \ x$.

Now, $\exp(-\alpha_k) \le \exp(- \sqrt{2 \ln(2)} x)$ and so $(1 - \exp(-\alpha_k))^{-1} \le (1-\exp(- \sqrt{2 \ln(2)} x))^{-1}$ for $k=1, \cdots, 2K$. The coefficients $\xi_k$ can be loosely bounded using the following upper bounds for binomial coefficients
\begin{eqnarray*}
{n \choose m} \le \left(\frac{n e}{m}\right)^m, \ \ \textrm{and} \ \ {n \choose m} \le \frac{n^m}{m!}.
\end{eqnarray*}
For $k=1, \cdots, 2K$, we have
\begin{eqnarray*}
\vert \xi_k \vert & \le & \frac{1}{K!} \sum_{j= \lfloor (k+1)/2 \rfloor}^{k \wedge K} j^{K+1} \left(\frac{Ke}{j}\right)^K (2e)^j j^k \\
                 & \le & \frac{1}{K!} \sum_{j= \lfloor (k+1)/2 \rfloor}^{k \wedge K} k^{k+1}(Ke)^K (2e)^k \\
                 & \le & \frac{1}{K!} \frac{k}{2} K^{K+2}(2e^2)^K
\end{eqnarray*}
so that
\begin{eqnarray*}
\sum_{k=1}^{2K} \frac{\vert \xi_k \vert}{k} & \le &  \frac{1}{K!}K^{2K+1} (2e^2)^K.
\end{eqnarray*}
Hence, if we impose that $\left \vert \tilde{E}_{K,N}(x) \right \vert < \epsilon$, then it is enough to choose $N$ such that
\begin{eqnarray*}
N > \frac{1}{\sqrt{2 \ln(2)} \ x } \left\{ \ln\left(\frac{1}{\epsilon (1 - \exp(- \sqrt{2 \ln(2)} \ x ))}\right) + (2K+1)\ln(K) + 3K+2 \right \}. \hspace{0.5cm} \Box
\end{eqnarray*}

\medskip

\par \noindent \textbf{Lemma A.2} \ \textit{For all $x > 0$, we have
\begin{eqnarray*}
x K_1(x) - K_0(x) \le \sqrt{\frac{\pi}{2}} \sqrt{x} \exp(-x).
\end{eqnarray*}
}

\medskip

\par \noindent \textbf{Proof.} \ Let us recall some well-known facts about modified Bessel functions of the second kind.
\begin{eqnarray}\label{PropK}
K_{1/2}(z) &=& \sqrt{\frac{\pi}{2}}  \frac{\exp(-z)}{\sqrt{z}}, \ \ \textrm{for all $z \in \mathbb C^*$} \label{PropK1}\\
K_n(z) & \approx & \sqrt{\frac{\pi}{2}}  \frac{\exp(-z)}{\sqrt{z}}, \ \textrm{as $\vert z \vert \to \infty$ and $n \in \mathbb N$} \label{PropK2}\\
\lim_{x \searrow 0} K_n(x) & = & \infty  \ \ \textrm{for all $n \in \mathbb N$} \label{PropK3}\\
K_1(z) & \approx & \frac{1}{z} \ \ \textrm{as $\vert z \vert \searrow 0$} \label{PropK4} \\
(z^n K_n(z))' &=& - z^n K_{n-1}(z), \ \ \textrm{for all $z \in \mathbb C$ and $n \in \mathbb Z$} \label{PropK5}\\
K'_n(z) & =& - \frac{n}{z} K_n(z) - K_{n+1}(z) \ \ \textrm{for all $z \in \mathbb C^*$ and $n \in \mathbb Z$} \label{PropK6}\\
K_\nu(x) & \le & K_{\nu'}(x) \ \  \textrm{for all $x > 0$ and $\nu < \nu' \in  \mathbb R$}. \label{PropK7}
\end{eqnarray}
see e.g. Abramowitz and Stegun 1964. Note first that by (\ref{PropK1}), the inequality stated in the lemma is equivalent to
 \begin{eqnarray*}
 xK_1(x) - K_0(x) \le x K_{1/2}(x), \ \ x > 0.
 \end{eqnarray*}
From (\ref{PropK1}), (\ref{PropK2}) and (\ref{PropK3}), it follows that
\begin{eqnarray*}
\lim_{x \searrow 0} (xK_1(x) - K_0(x) - xK_{1/2}(x)) &= &- \infty \ \ \textrm{and} \\
\lim_{x \to \infty} (xK_1(x) - K_0(x) - xK_{1/2}(x)) &= &0.
\end{eqnarray*}
Let us write $\psi(x) = xK_1(x) - K_0(x) - xK_{1/2}(x), x > 0$. Suppose now that there exists $x > 0$ such that $\psi(x) > 0$. This would imply that there exists $y > 0$ such that $\psi(y) > 0$ and $\psi'(y) =0$. Now, using (\ref{PropK5}) and (\ref{PropK6}) it follows that
\begin{eqnarray*}
\psi'(x) = -xK_0(x) + K_1(x) + \left(x - \frac{1}{2}\right) K_{1/2}(x), \ x > 0.
\end{eqnarray*}
Hence, $y$ satisfies
\begin{eqnarray*}
y K_1(y) - K_0(y) & > &  y K_{1/2}(y) \ \ \textrm{and} \\
K_1(y) & =  &\left(\frac{1}{2} - y\right) K_{1/2}(y) + y K_0(y).
\end{eqnarray*}
It follows that
\begin{eqnarray*}
(y^2 - 1) K_0(y) >  y \left(y + \frac{1}{2} \right) K_{1/2}(y).
\end{eqnarray*}
Since $K_0(x) > 0$ and $K_{1/2}(x) > 0$ for all $x > 0$, we must have $y > 1$. But if $y > 1$, then the previous inequality implies
\begin{eqnarray*}
 K_0(y) >  \frac{y (y + 1/2)}{y^2 - 1} K_{1/2}(y) > K_{1/2}(y)
\end{eqnarray*}
which is impossible by (\ref{PropK7}).

\end{document}